 \documentclass[numberedappendix,twocolumn,twocolappendix]{openjournal}

\usepackage{amssymb,amsmath}
\usepackage{lipsum}
\usepackage{orcidlink}
\usepackage{natbib}

\usepackage{xcolor}
\usepackage{hyperref}
\hypersetup{
    unicode, 
    colorlinks=true,
    linkcolor=linkcolor,
    citecolor=linkcolor,
    filecolor=linkcolor,
    urlcolor=linkcolor,
}
\definecolor{linkcolor}{rgb}{0.0,0.3,0.5}

\newcommand{\pd}{\partial}

\newcommand{\mcF}{\mathcal{F}}

\newcommand{\mcL}{\mathcal{L}}

\newcommand{\mcO}{\mathcal{O}}

\newcommand\vecd{\mathbf{d}}

\newcommand\vecf{\mathbf{f}}

\newcommand\vecm{\mathbf{m}}

\newcommand\vecq{\mathbf{q}}

\newcommand\vecs{\mathbf{s}}
\newcommand\vect{\mathbf{t}}
\newcommand\vecu{\mathbf{u}}

\newcommand\vecw{\mathbf{w}}
\newcommand\vecx{\mathbf{x}}
\newcommand\vecy{\mathbf{y}}
\newcommand\vecz{\mathbf{z}}

\newcommand\vecD{\mathbf{D}}

\graphicspath{ {./} }

\begin{document}

\title{Generating Dark Matter Subhalo Populations Using Normalizing Flows}


\author{Jack Lonergan\orcidlink{0009-0000-5753-8918}}
\email{jacklone@usc.edu}
\affiliation{Department of Physics and Astronomy, University of Southern California}

\author{Andrew Benson\orcidlink{0000-0001-5501-6008}}
\email{abenson@carnegiescience.edu}
\affiliation{The Observatories of the Carnegie Institution for Science}

\author{Daniel Gilman\orcidlink{0000-0002-5116-7287}}
\email{gilmanda@uchicago.edu}
\affiliation{Department of Astronomy and Astrophysics, University of Chicago}

\begin{abstract}
    Strong gravitational lensing is a powerful tool for probing the nature of dark matter, as lensing signals are sensitive to the dark matter substructure within the lensing galaxy. We present a comparative analysis of strong gravitational lensing signatures generated by dark matter subhalo populations using two different approaches. The first approach models subhalos using an empirical model, while the second employs the {\sc Galacticus} semi-analytic model of subhalo evolution. To date, only empirical approaches have been practical in the analysis of lensing systems, as incorporating fully physical models was computationally infeasible. To circumvent this, we utilize a generative machine learning algorithm, known as a normalizing flow, to learn and reproduce the subhalo populations generated by {\sc Galacticus}. We demonstrate that the normalizing flow algorithm accurately reproduces the {\sc Galacticus} subhalo distribution while significantly reducing computation time compared to direct simulation. Moreover, we find that subhalo populations from {\sc Galacticus} produce comparable results to the empirical model in replicating observed lensing signals under the fiducial dark matter model. This work highlights the potential of machine learning techniques in accelerating astrophysical simulations and improving model comparisons of dark matter properties.
\end{abstract}









\maketitle




\section{Introduction}
\label{introduction}

Dark matter is an unknown form of non-baryonic matter that composes up to $85\% $ of all matter in the universe \citep{ade2016planck, aghanim2020planck}. Given that dark matter constitutes the majority of matter in the universe, it plays a significant role in the formation and evolution of structure throughout the universe's history. Quantum fluctuations in the early universe produced slight overdensities in an otherwise homogeneous matter density field, and these overdensities eventually formed gravitationally bound structures known as dark matter halos \citep{spergel2003first, hinshaw2013nine}. These halos merged over time, resulting in a present-day hierarchical distribution of matter \citep{1985ApJ...292..371D,cole1996structure, makishima1998hierarchical}. The current cosmological model is the Lambda cold dark matter ($\Lambda$CDM) which accurately models the large-scale structure of the universe \citep{frenk1985cold, peebles2020principles}. The fiducial model begins to experience tensions with observations on smaller, galactic scales, however, and many studies have proposed models to accurately replicate the distribution of dark matter to match observations on these scales \citep[see][for a review of problems and proposed solutions]{bullock2017small}.

A wide range of techniques for simulating the formation and evolution of dark matter substructure in galaxies have been developed, ranging from analytic models \citep{2005ApJ...624..505Z,2008MNRAS.387..689G,2011ApJ...741...13Y} to N-body simulations \citep{1999ApJ...524L..19M,2008MNRAS.391.1685S, nadler2023symphony}. Analytic models offer an efficient way to model a broad range of halos at the cost of making simplifying physical assumptions to allow the model to be solved analytically. N-body simulations, on the other hand, numerically solve the equations governing an evolving matter density field in an expanding universe. The development of such simulations has allowed us to replicate the distribution of matter at high resolutions over large time intervals, providing a detailed picture of the evolution of the universe \citep{mikkola1993implementation, bertschinger1998simulations, bagla2005cosmological}. Solving these equations numerically, however, demands significant computational resources, making the computational cost one of the main drawbacks of N-body simulations \citep{dehnen2011n}. One middle-ground approach between analytic models and simulations is the class of semi-analytic models (SAMs) of subhalo evolution. As with $ N $-body simulations, these models perform simulations with the modification that, in SAMs, the fundamental objects are halos themselves, rather than individual mass particles. These halos take on simple geometries (i.e., spherical halos), which allows for some solutions to become analytic and greatly reduces the computational cost where analytic solutions are unavailable \citep{2001ApJ...559..716T,2004MNRAS.351.1215B,2005ApJ...624..505Z, pullen2014nonlinear,2022PhRvD.106l3026D}. Each method presents a trade-off between computational cost and accuracy, so different models are chosen depending on the specific requirements of their application.

In addition to a wide variety of approaches to modeling dark matter, there are also many ways in which we can detect dark matter, despite its lack of interactions with electromagnetic radiation. Here, we specifically focus on one such method---strong gravitational lensing of quasar images---which can be used to constrain dark matter particle properties \citep{gilman2020warm, vegetti2014density, hezaveh2016detection}. As light passes by a galaxy and its dark matter halo, its path is deflected by the gravitational potential of that galaxy, its halo, and the population of subhalos. The resulting distortion of the quasar image can be quantified to constrain the underlying population of subhalos, their distribution of masses, spatial distribution, and density profiles. While strong gravitational lensing of galaxies is merely one method for constraining dark matter models, it is considered advantageous over other methods such as analyzing the Lyman-$ \alpha $ forest or counting of galaxies because the effect of gravitational lensing only depends on the total gravitational potential, which, in the low mass (sub)halos to which this method is sensitive, is dominated by dark matter and is not expected to be significantly altered by baryonic processes \citep{2016MNRAS.457L..74D}. As such, strong gravitational lensing bypasses the need to understand the complex, baryonic physics of the system \citep{treu2010strong,gannon2025dark}. Additionally, there are typically high degrees of uncertainty when attempting to distinguish the gravitational effects of dark matter when also accounting for baryonic matter \citep{2019MNRAS.490..962F}, making it convenient to work with measurements that are largely unaffected by baryonic physics.

Previous studies have shown that dark matter models can be constrained using imaging of gravitationally strongly lensed quasars \citep{dalal2002direct, nierenberg2014detection, xu2015well, hsueh2020sharp, gilman2021strong, nadler2021dark}. For example, \cite{gilman2020warm} used a parameterized model to generate dark matter halo populations within the lensing galaxy (which going forward will be referred to as the ``empirical model''), repeating this process of creating halo populations between 300,000 to 1,200,000 times for a given lens system. This was done to obtain a statistically representative distribution of subhalo populations responsible for the lensing signature and thereby assess the relative likelihood of a given dark matter model reproducing the observed lensing statistics. While this model accurately reproduces certain summary statistics of the subhalo population, such as the subhalo mass function and distribution of infall redshifts, it also makes some substantial simplifying assumptions. Specifically, it does not fully account for correlations between
the orbital properties of a subhalo and tidal stripping (instead it connects tidal stripping to only the current orbital radius), growth of the host halo, and the radial distribution of subhalos. Work has been done to account for the effects of tidal stripping in analytic models \citep[e.g.][]{han2016unified}, yet underlying physical assumptions (such as a mass-independent tidal stripping function) remain present. The impact of such assumptions made in analytic models could be examined if halo populations used in the analysis accounted for these correlations (i.e. halo populations from SAM's or N-body simulations). The primary obstacle preventing such populations from being implemented is that it is computationally infeasible to generate $ \mcO(10^6) $ realizations of dark matter halos directly from either SAMs or N-body simulations.

To bypass these limitations, we propose a more computationally efficient method to produce large numbers of realizations of subhalo populations with statistical properties as predicted from simulations. Specifically, we make use of a generative machine learning algorithm known as a normalizing flow. We train the normalizing flow on a set of simulation data, and then, once trained, we can sample new subhalo populations in a fraction of the time it would take to generate populations of subhalos directly from simulations. In principle, this methodology can be applied to any relevant simulation data set, but here we make use of simulated data from the {\sc Galacticus}\footnote{\href{https://github.com/galacticusorg/galacticus}{https://github.com/galacticusorg/galacticus}, we use revision \href{https://github.com/galacticusorg/galacticus/commit/4313a4e2c7b68abbec547107e118ad22b33fc556}{4313a4e2c7b68abbec547107e118ad22b33fc556}.} semi-analytic model \citep{benson2012galacticus} which has also been used to guide the construction of the empirical model in \cite{gilman2020warm}. In addition to generating {\sc Galacticus} halo populations using the normalizing flows algorithm, we show an example application in which we apply the generated subhalo realizations to the forward modeling approach of \cite{gilman2020warm} for a single lensed system.

This paper is organized as follows: In section \ref{sec:methods}, we introduce the empirical and {\sc Galacticus} halo population models, as well as the normalizing flows algorithm. In section \ref{sec:constrainingDM}, we detail the lensing analysis conducted in \citep{gilman2020warm} and describe the implementation of emulator subhalo populations into this analysis. In section \ref{sec:r&a}, we cover the emulator's efficiency and present the results of the forward modeling application. In section \ref{sec:limitations}, we discuss the limitations of this machine learning approach and possibilities for future work. Finally, section \ref{sec:conclusions} presents our general conclusions.

\section{Methods} \label{sec:methods}

In this section, we first describe two different models for subhalo populations. We summarize the empirical model and then describe in detail the properties of {\sc Galacticus} and how it generates a realization of a subhalo population. We then describe the specifics of the normalizing flows based algorithm used to replicate {\sc Galacticus} realizations

\subsection{Models}

\subsubsection{The Empirical Model}

The dark matter halos in the empirical model are constructed from a series of analytic models---describing the number, spatial distribution, and density profiles---for both the cold dark matter and warm dark matter models, and for both subhalos and halos along the line of sight of the lensing galaxy. This paper, however, will focus specifically on CDM subhalo populations. The reason for this is that the properties of line-of-sight halos are relatively simple as they do not undergo the strong tidal evolution experienced by subhalos and already have accurately fitting equations \citep{2002MNRAS.329...61S,navarro1996structure}. Subhalos, on the other hand, are influenced by complicated, non-linear evolutionary processes such as tidal stripping, tidal heating, tidal truncation, dynamical friction, etc., whose intricate details are not fully accounted for in the empirical model. So it is these halos whose properties will be most impacted by analytic approximations. The details of the analytic model can be found in \cite{gilman2020warm}, with tidal physics updated as described in \cite{2024MNRAS.535.1652K}, but we summarize here the main points relevant to CDM subhalos:

\begin{itemize}
    \item Subhalo density profiles are modeled as truncated Navarro-Frenk-White \citep[NFW][]{navarro1996structure} profiles:
\end{itemize}

\begin{equation}
    \rho(r) = \frac{\rho_{\mathrm s}}{x(1 + x)^2} \frac{\tau^2}{x^2 + \tau^2},
\end{equation}
where $ x = r/r_{\mathrm s} $, $ \tau = r/r_{\mathrm t} $, $ \rho_{\mathrm s} $ is the scale density, $ r_{\mathrm s} $ is the scale radius, and $ r_{\mathrm t} $ is the truncation radius. Halo masses are defined as the total mass with a spherical region of radius $r_{200}$ within which the mean density is 200 times the critical density of the universe at the current redshift. 

\begin{itemize}
    \item The truncation radius of a given subhalo is defined by $ r_t = r_{50} $, where $ r_{50} $ denotes the radius which encloses 50 times the critical density \citep{keeley2024jwst}.

    \item Subhalos in the empirical model are spatially distributed in 3D inside the virial radius of the host halo. Within this volume, subhalos are assumed to trace the mass profile of the host outside of an inner radius which is set equal to half the host scale radius. Inside that inner radius, subhalos are distributed uniformly in three dimensions, motivated by tidal disruption simulations around lens galaxies \citep{jiang2017statistics}. This results in a distribution that is approximately uniform in projection in the inner region of the halo that is relevant to lensing, in agreement with N-body simulations \citep{xu2015well}.
    \item The subhalo mass function is given by:
\end{itemize}
\begin{equation}
    \frac{\mathrm{d}^2 N_{\text{sub}}}{\mathrm{d}m\mathrm{d}A} = \frac{\Sigma_\text{sub}}{m_0} \left( \frac{m}{m_0} \right)^\alpha \mcF(M_\text{halo}, z),
\end{equation}
where $ \Sigma_\text{sub} $ is a normalization parameter characterizing the amplitude of the mass function, the pivot is denoted by $ m_0 = 10^8 \mathrm{M}_\odot $, and $ \mcF(M_\text{halo}, z) $ accounts for how the number density of halos depends on both the host halo mass and redshift \citep{gannon2025dark}.  

\subsubsection{{\sc Galacticus} Model}

The second model that we use to generate subhalo populations in this work is the semi-analytic galaxy formation model known as {\sc Galacticus} \citep{benson2012galacticus}. {\sc Galacticus} is a highly modular code that simulates the evolution of galaxies and their dark matter halos by representing them as individual nodes in an overall merger tree. As these nodes evolve, they undergo merging events resulting in the hierarchical assembly of dark matter halos. Constructing a set of evolved dark matter halos in {\sc Galacticus} requires two main steps: First, a merger tree is constructed, which represents, at each point in time, the set of progenitor halos from which the final halo is formed. Then, each node is evolved forward in time by numerically solving a series of differential equations that describe the relevant physics.

The initial step in setting up a halo population is to generate the merger tree. This is done using a Monte Carlo (MC) method to match predictions from the extended Press-Schechter (EPS) formalism \citep{1991MNRAS.248..332B,1994MNRAS.271..676L}. The details of how {\sc Galacticus} merger trees are constructed can be found in \cite{cole2000hierarchical}, with merger rates updated to match the calibrated results of \cite{parkinson2008generating}. These calibrations were shown to be in good agreement with the measurements of progenitor mass functions from N-body simulations. For this study, merger trees were simulated for a host halo of mass $M_\mathrm{lens} = 10^{13.3} \mathrm{M}_\odot $ at $ z_\mathrm{lens} = 0.5 $ with a mass resolution of $ 10^6 \mathrm{M}_\odot $. This host mass and lens redshift were chosen to match the conditions of a typical strong lens galaxy. The mass resolution was chosen to be $ 10^6 \mathrm{M}_\odot $ as halos below this mass have negligible contributions to the overall lensing signal. As discussed by \cite{gilman2020warm}, the lensing signal of a subhalo becomes undetectable somewhere in the mass range $ 10^6$--$10^7 \mathrm{M}_\odot $, so $ 10^6 \mathrm{M}_\odot $ acts as a conservative lower limit. Subhalos were also restricted to lie within a 20 kpc annulus from the center of the host, as it is subhalos within the vicinity of the Einstein radius that primarily contribute to the observed lensing signature \citep{gannon2025dark}.

Once a merger tree is constructed in {\sc Galacticus}, halos in the tree are evolved forward in time by numerically solving differential equations to produce a population of dark matter halos at the epoch of interest (in this case, the redshift of the lens galaxy). Specifically, {\sc Galacticus} tracks the evolution of the total mass of a given halo following the mass history along each branch of the merger tree. Orbital properties of subhalos, such as their 3D position, velocity, bound mass, and density profile, are tracked by incorporating the effects of the host halo potential, along with dynamical friction, tidal stripping, and tidal heating \citep{pullen2014nonlinear, yang2020new, benson2022tidal}. {\sc Galacticus} also computes the evolution of the dark matter profile scale length, determined from the halo's concentration \citep{diemer2019accurate}.

When generating subhalo realizations using {\sc Galacticus} directly, numerically solving differential equations for the evolution of each halo takes substantial amounts of time. In this work, we generated 300 merger trees in total, with each individual tree taking between 6--8 CPU-hours of computation time to evolve to $z_\mathrm{lens}$. It would be computationally impractical to generate $ \mcO(10^6) $ of these trees for each lens to be studied, which motivates the search for more computationally efficient alternatives.

\subsection{The Normalizing Flows Algorithm}

While {\sc Galacticus} is capable of simulating populations of dark matter subhalos directly, we instead propose training a normalizing flow emulator on data generated by {\sc Galacticus}, and then using that emulator to generate populations of subhalos statistically consistent with those produced by {\sc Galacticus}. Such algorithms take in an input data distribution and learn a mapping from a simple, latent space to the input data space. The mapping between latent and data spaces has some convenient transformation properties such as being invertible and at least once differentiable. The invertible mapping allows the flow-based algorithm to perform both sampling and likelihood estimation \citep{papamakarios2021normalizing}.

The flows-based algorithm used in this work takes in a data distribution $ p(\vecz) $ and generates an invertible map $ F $ between the data distribution and a latent distribution $ \tilde{p}(\vecu) $. The relationship between the prior and likelihood distributions is modeled using the change of variables formula:

\begin{equation}
    p(\vecz) = \tilde{p} \left( F^{-1}(\vecz) \right) \left| \det \left( \frac{\pd F^{-1}}{\pd \vecz} \right) \right|.
\end{equation}

Here, the final term is the determinant of the Jacobian of $ F^{-1} $ evaluated at each component of $ \vecz $. This term is included to account for the change in volume when transforming between the latent and data spaces. The mapping $ F $ is composed of $ n $ individual transformations $ F = f_1 \circ f_2 \circ \dots \circ f_n $, each of which is independent, differentiable, and invertible. As the data passes through the $ k^\mathrm{th} $ individual transformation in the mapping $ f_k $, it satisfies $ \vecw_k = f_k(\vecw_{k - 1}) $. Here, $ \vecw_k $ represents the data at stage $ k $ of the transformation, so $ \vecw_0 = \vecu $ and $ \vecw_n = \vecz $. 

In this work, our normalizing flow is composed of $ n = 12 $ affine coupling layers, which are individual bijective functions that construct the map. An individual affine transformation $ f $ acting on the data $ \vecw_k $ is defined by
\begin{equation}
    f(\vecw_k; \vecs, \vect) = \exp(\vecs) \odot \vecw_k + \vect,
\end{equation}
where $ \odot $ represents element-wise multiplication, and $ \vecs $ and $ \vect $ are the vectors that contain the scale and translation coefficients for the affine transformation. The $ \vecs, \vect $ coefficient vectors are obtained by passing previous iterations of the data $ \vecw_{1:k - 1} $ through a neural network of 5 dense layers: four intermediate ReLU activation functions, followed by a tanh activation function. Using 5 dense layers to generate the scaling and translation coefficients of the affine transformation adds complexity and non-linearity to the model. To prevent overfitting, each individual scaling and transformation layer has an L2 regularization metric with a regularization parameter of 0.01. The overall model is compiled using the Adam optimizer with a learning rate of 0.0001 \citep{loshchilov2017decoupled}.

The normalizing flows algorithm is specifically applied to an input data distribution that parametrizes {\sc Galacticus} subhalos. Each subhalo is characterized by an infall\footnote{``Infall'' here means the time at which a halo first became a subhalo, by falling inside the virial radius of a larger halo.} mass $ M_\text{infall} $, concentration $ c $, bound mass $ M_\text{bound} $, infall redshift $ z_\text{infall} $, truncation radius $ r_{\mathrm t} $ and two dimensional projected radius\footnote{For the purposes of gravitational lensing, only the projected radius is relevant as all subhalos effectively lie in plane---the extent of the subhalo population along the line of sight (of order the virial radius of the host halo) is negligible compared to the distance from observer to lens, and lens to source.} $ r_\text{2D} $ taken from {\sc Galacticus} realizations where these parameters were computed by solving the differential equations governing subhalo evolution, and together fully characterize the properties of a subhalo that are needed for lensing calculations. The collection of these six parameters for each generated subhalo forms a six-dimensional parameter space for the population of subhalos. Before this 6D distribution is input into the flows algorithm, the un-normalized data $ \vecx = (M_\text{infall}, c, M_\text{bound}, z_\text{infall}, r_{\mathrm t}, r_{2D}) $ is first normalized to a new vector $ \vecy $ whose components are defined through the following set of equations:
\begin{align}
    y_1 &= \log_{10}(M_\text{infall}/M_\text{host}), \hspace{3ex} \\
    y_2 &= c, \\
    y_3 &= \log_{10}(M_\text{bound}/M_\text{infall}), \\
    y_4 &= z_\text{infall}, \\
    y_5 &= \log_{10}(r_{\mathrm t}/R_\text{vir, host}), \\
    y_6 &= \log_{10}(r_\text{2D}/R_\text{vir, host}).
\end{align}
In these equations, the variables $ M_\text{host} $ and $ R_\text{vir, host} $ refer to the virial mass and radius, respectively, of the host halo as computed from {\sc Galacticus}. These virial quantities are defined in terms of the virial density contrast $ \Delta_\text{vir} $, which in this work is based on the spherical collapse model in a universe that contains collisionless matter and a cosmological constant \citep{bryan1998statistical}. Only subhalos along the line of sight within a double cone with opening angle $ \theta = 3R_\textrm{Ein} $, where $ R_\textrm{Ein} $ is the Einstein radius, contribute to the lensing signal \citep{gilman2020warm}. 

To evaluate the truncation radii of  {\sc Galacticus} subhalos, the ratio $ \rho(r)/\rho_\text{NFW}(r) $ of a given subhalo is evaluated at a range of radii ranging from the innermost regions of the halo out to the virial radius. Here, $ \rho(r) $ is the actual density profile generated by {\sc Galacticus} including the effects of tidal stripping and heating, and $ \rho_\text{NFW}(r) $ is the NFW density profile of the halo prior to any tidal effects. A truncation fraction $ f_\text{trunc}(r, r_{\mathrm t}) $ is then defined in terms of the truncation radius, $r_{\mathrm t}$:
\begin{equation}
    f_\text{trunc}(r,r_{\mathrm t}) = \frac{1}{1 + (r/r_{\mathrm t})^2}.
\end{equation}
To find the best-fit truncation radius we minimize the objective function $ \chi^2(r_{\mathrm t}) $:
\begin{equation}
    \chi^2(r_{\mathrm t}) = \sum_{r \leq R_\text{vir}} \left( \log_{10} \frac{\rho(r)/\rho_\text{NFW}(r)}{f_\text{trunc}(r, r_{\mathrm t})} \right)^2.
\end{equation}

Once the {\sc Galacticus} data has been normalized, it is then subsequently shifted so that it lies within a hypercube of side length 2, centered at the origin, with each coordinate in the range $ [-1, 1] $. This newly transformed data will be referred to as the hypercube data $ \vecz $, and its components are defined in terms of the normalized data $ \vecy $ through the following definition:
\begin{equation} \label{hypercube}
    z_i = 2 \sigma_{i} - 1, \hspace{3ex} \sigma_{i} = \frac{y_i - y_{\text{min}, i}}{y_{\text{max}, i} - y_{\text{min}, i}}.
\end{equation}
Here, $ y_i $ refers to the $ i^\mathrm{th} $ component of the normalized data, and $ y_{\text{min}, i}, y_{\text{max}, i} $ are the minimum and maximum values of the $ i^\mathrm{th} $ parameter, respectively. It should be noted that equation \ref{hypercube} is invertible, so it is straightforward to transform the data out of the hypercube coordinates back to their starting, normalized values. It is this hypercube data, $\vecz$, that is fed into the emulator as input data. The purpose of this additional shift from $ \vecy $ to $ \vecz $ is so that data points do not exponentially grow as they pass through layers of the emulator, leading to potential numerical instability. Keeping the data restricted to a $ [-1, 1] $ interval makes it easy to transform as it propagates through the neural network. From the total input data, an 80/20 split is implemented between a validation and training set for the model.

\begin{figure*}
    \includegraphics[width = \textwidth]{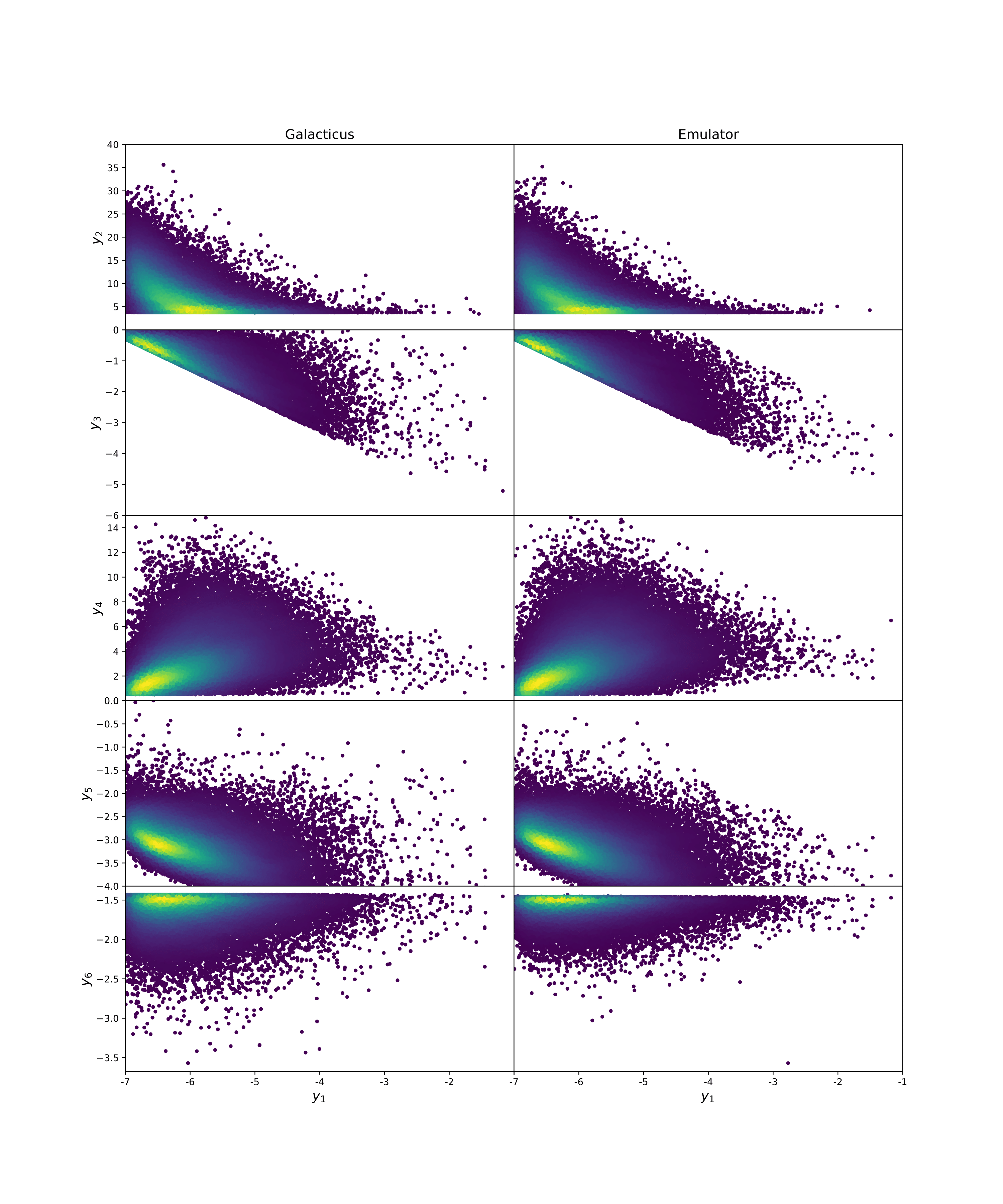}
    \centering
    \caption{Two dimensional density distributions of normalized subhalo parameter spaces generated from {\sc Galacticus} (left column) versus the emulator (right column). Yellow regions correspond to regions of higher number density. The parameters $ y_1, \dots, y_6 $ denote the normalized infall masses, concentrations, bound masses, infall redshifts, truncation radii, and projected radii respectively.}
    \label{density}
\end{figure*}

In Figure \ref{density}, we show the results for the emulator's subhalo population in comparison to the population from {\sc Galacticus}. The left column shows normalized 2D density plots of {\sc Galacticus} subhalos. The right column is the emulator's attempt to replicate the distribution. Both columns show a total of 49,633 subhalos sampled from their respective distributions. Overall, the similarities between the two columns indicate that the emulator can accurately learn the normalized {\sc Galacticus} subhalo distribution. A more detailed discussion quantifying the accuracy of the emulator is in Appendix \ref{sec:appA}. It should be noted that the emulator's invertible mapping, plus the fact that the emulator knows nothing about the physics of subhalos (it just views the input as a distribution of data points), will have some difficulties modeling sharp edges at the boundaries of the distribution. Therefore, the emulator plots are clipped to ensure that each of the subhalos obeys the following physical and model constraints:

\begin{itemize}
    \item Subhalos must have infall masses greater than twice the mass resolution, specifically $ M_\text{infall} > 2 \times 10^6 \mathrm{M}_\odot $. This constraint is a result of how {\sc Galacticus} generates halos in a merger tree. A given halo will have two progenitor halos of smaller masses, and a halo with $ M_\text{infall} < 2 \times 10^6 \mathrm{M}_\odot $ would necessarily imply that at least one of its progenitors is below the mass resolution. Since subhalos below the mass resolution (by definition) are not included, {\sc Galacticus}'s population of (sub)halos will start to become incomplete below this limit.
    \item Emulator subhalos must have bound masses within the range  $ [10^6, 10^{9}] \ \mathrm{M}_\odot $. The lower bound is enforced as, during the evolution of subhalos in {\sc Galacticus}, if the bound mass falls below $10^6\mathrm{M}_\odot$, the subhalo is removed from the calculation. The upper bound is imposed to match the range of subhalo masses considered in the empirical model. 
    \item The bound masses of a subhalo must be less than its corresponding infall mass, since, in {\sc Galacticus}, subhalos can only lose mass after infall due to tidal stripping.
    \item The infall redshifts of emulated subhalos must be greater than or equal to the redshift of the lensing galaxy, specifically $ z_\text{lens} \geq 0.5 $---by definition, for a halo to be a subhalo by the time it is observed at $z_\mathrm{lens}$ it must have fallen in at some earlier time, $z_\mathrm{infall} > z_\mathrm{lens}$. 
    \item We restrict ourselves to looking at subhalos in a narrow region around the Einstein radius, as these are the subhalos that predominantly contribute to the strong gravitational lensing signal. Specifically, this is enforced by requiring $ r_\text{2D} \leq 20 $ kpc. A similar choice is made in the empirical model.
\end{itemize}

\section{Constraining Dark Matter from the Lensing Signature} \label{sec:constrainingDM}

We have shown that the normalizing flows algorithm is able to accurately replicate the distribution of {\sc Galacticus} dark matter subhalos. The power of the normalizing flows algorithm, however, is not limited to its ability to reproduce subhalo populations. The broader motivation for implementing a flows-based algorithm is that the emulated subhalos, with the same physical accuracy as {\sc Galacticus} subhalos, can be efficiently implemented into any analysis that utilizes subhalo populations.

In this section, we describe how subhalos emulated from {\sc Galacticus} can be implemented into an analysis code that measures gravitational lensing signatures. Specifically, emulator subhalos are input into the code {\sc pyHalo}\footnote{\href{https://github.com/dangilman/pyHalo}{https://github.com/dangilman/pyHalo}, we use revision \href{https://github.com/dangilman/pyHalo/commit/9a5d619f3feef297fce8b0a302862991dcb2bd74}{9a5d619f3feef297fce8b0a302862991dcb2bd74}.} which outputs summary statistics describing how well input halo realizations match observational data from observed lensed quasar systems \citep{gilman2024turbocharging}. We extended {\sc pyHalo} to support subhalos with the emulator's parameterization, which differs in detail from that of the empirical model. A more detailed description of the analysis process is depicted in \cite{gilman2020warm}, but we will outline the general analysis process here.

The overall goal is to utilize a forward-generative model that samples the target posterior distribution using an Approximate Bayesian Computing (ABC) technique. Specifically, the posterior distribution of interest is the distribution of subhalo population parameters $ \vecq_\mathrm{s} $ given the data $ \vecD $, which comes in the form of quasar image positions and flux ratios. Mathematically, this distribution can be modeled by the relation:
\begin{equation}
    p(\vecq_\mathrm{s} | \vecD) \propto \pi(\vecq_\mathrm{s}) \prod_{n = 1}^N \mcL(\vecd_n|\vecq_\mathrm{s}).
    \label{posterior}
\end{equation}
Here, $ N $ refers to the number of lenses in the system, $ \pi(\vecq_\mathrm{s}) $ is the prior distribution of the dark matter parameters, and $ \vecd_n $ is the data from the $ n^\mathrm{th} $ lens. In this work, the target data we use is the set of image positions and flux ratios for a smooth simulated mock lens from an elliptical power law mass profile with Einstein radius of 1 arc second, axis ratio of 0.73, a logarithmic power law slope of 2.05, and an external shear strength of 0.057. The term ``smooth'' here refers to a lens system with no substructure. While equation \ref{posterior} provides a straightforward approach to posterior sampling in theory, the likelihood is difficult to evaluate in practice due to its large dimensionality. Therefore, an alternative approach can be taken where the data itself is forward modeled.

Instead of directly sampling the likelihood, an alternative approach is to first create simulated data which has lenses $ \vecd'_n $. This data set is composed of quasar image positions and flux ratios produced from a simulated realization of subhalos $ \vecm_\text{sub} $ generated from model parameters $ \vecq_\mathrm{s} $. Then, in the case where $ \vecd_n = \vecd'_n $ for some $ n $, we will have found values for $ \vecq_\mathrm{s} $ which have correctly reproduced the observational data and thus have found a single data point in the posterior distribution. By generating many different subhalo realizations $ \vecm_\text{sub} $ for different $ \vecq_\mathrm{s} $ and finding which realizations most accurately reproduce the data, we can effectively sample from the posterior distribution without needing to calculate the likelihood directly.

To quantify how well a given realization of subhalos reproduces observed data, a summary statistic $ S_\text{lens} $ as follows:
\begin{equation}
    S_\text{lens}(\vecf, \vecf') \equiv \sqrt{\sum_{i = 1}^3 \left( f'_i - f_{\text{obs}(i)} \right)^2}.
\end{equation}
The vector $ \vecf $ represents the flux ratios from the target data, and the vector $ \vecf' $ represents flux ratios produced from the set of simulated subhalos. Given that the object being observed is a quadruply imaged quasar, there will naturally be four lensed images and, thus, three flux ratios. The statistic $ S_\text{lens} $ represents a discrepancy between the observed and simulated flux ratios, such that the most accurate realizations will have the smallest values of $ S_\text{lens} $. A single $ S_\text{lens} $ value corresponds to a given subhalo population, and the distribution of $ S_\text{lens} $ values over many realizations provides insight into a model's relative ability to reproduce target data.

\section{Results and Analysis} \label{sec:r&a}

In this section, we present two sets of results from our study. First, we give a summary of run times needed to generate subhalo populations and perform lensing analysis. We then show the results of implementing emulator subhalos into the gravitational lensing analysis from the previous section.

\subsection{Summary of Run Times}

Table~\ref{subhalo_times} presents the run times needed to produce a subhalo population for the empirical model, direct generation from {\sc Galacticus}, and the emulator trained on {\sc Galacticus} data.

\begin{table}[hbt!]
    \centering
    \caption{Run times for generating a single subhalo population.}
        \begin{tabular}{ll}
            \hline
            \textbf{Model} & \textbf{CPU time [s]} \\
            \hline
            Empirical model & 2 \\
            {\sc Galacticus} & $2.6 \times 10^{4}$ \\
            Emulator & 2 \\
            \hline
        \end{tabular}
        \label{subhalo_times}
\end{table}

Each value measures the time taken to generate a single population of subhalos (for the specific host halo and resolutions described above). Here, the ``Empirical model'' row refers to the amount of time needed to generate a realization of subhalos from the empirical model, while the ``Galacticus'' row refers to how much time it takes to generate a realization of subhalos from {\sc Galacticus} directly. The ``Emulator'' row refers to how much time it takes the normalizing flows algorithm to generate a single realization of normalized subhalos (and subsequently un-normalize them) once trained. Although not mentioned in the table, it took around $1.8\times 10^4$~s to train the emulator on {\sc Galacticus} data. Of course, the emulator needs to be trained only once on an input data set, after which any number of subhalo populations with the same dark matter model, host halo mass, and redshift as the input data can be generated. 

We see that the time to train the emulator is less than the time it takes to generate a single subhalo population directly from {\sc Galacticus}, and note that the training dataset consisted of 300 {\sc Galacticus} subhalo population realizations. This highlights the computational efficiency of using the emulator to generate many populations of subhalos, as once the emulator is trained it takes only a couple of seconds to generate subsequent realizations. Moreover, we see that the time to generate realizations from the emulator (once trained) is similar to the time it takes to generate realizations from the empirical model\footnote{Furthermore, this time is small compared to the time taken to perform lensing calculations for a given subhalo population, such that subhalo population generation is not a significant contributor to the total time taken to analyze lensing data.}. Therefore, once the emulator learns the mapping between the latent and data spaces we can produce realizations of {\sc Galacticus} subhalo populations on timescales comparable to that of the empirical model.

Once subhalo populations are created either from either the empirical model or the emulator, it takes similar amounts of time for the forward modeling analysis to be applied to these realizations. This is to be expected since regardless of whether the empirical or emulator model is being applied, an $ N \times 6 $ array is fed into {\sc Samana}\footnote{\href{https://github.com/dangilman/samana}{https://github.com/dangilman/samana}, we use revision \href{https://github.com/dangilman/samana/commit/30cdf83609e50bdc4a0a7d89877cf613951b7115}{30cdf83609e50bdc4a0a7d89877cf613951b7115}.} (where $ N $ is the number of subhalos, each of which is described by 6 parameters). It takes around 80s to run {\sc Samana} and produce a single $ S_\text{lens} $ value. Therefore, producing 1,000,000 $ S_\text{lens} $ values takes $\sim$22,000 CPU-hours. Generating 300 {\sc Galacticus} trees and training the emulator takes only $\sim$2,100 CPU-hours. The process of generating training data and training the emulator is, therefore, not the bottleneck in the overall pipeline for producing $ S_\text{lens} $ distributions\footnote{Both {\sc Galacticus} and {\sc Samana} can be run in parallel, greatly reducing the wall-time needed to complete these calculations, but this does not change the overall number of CPU-hours required in each case.}.

\subsection{Forward Modeling the Data}

\begin{figure*}
    \includegraphics[width = \linewidth]{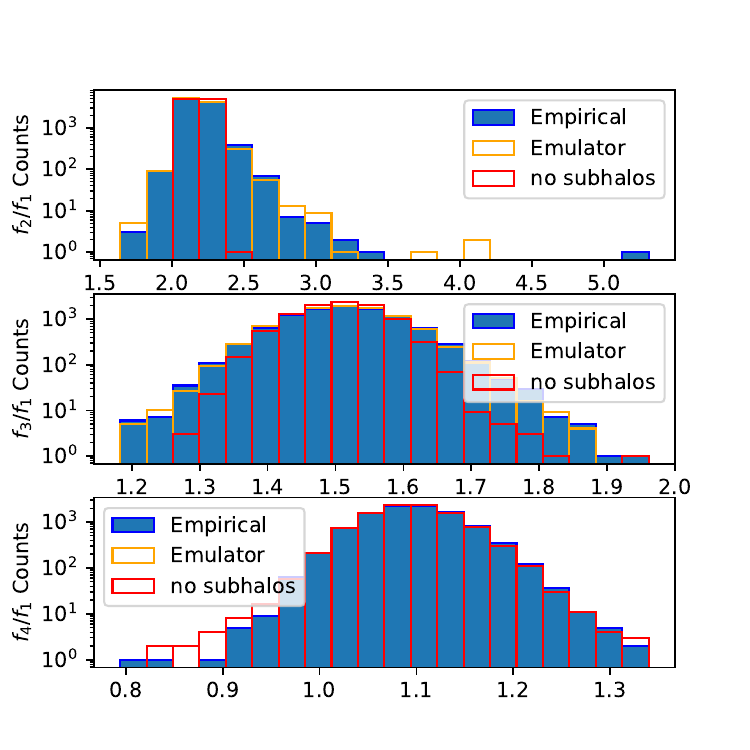}
    \centering
    \caption{Histograms for the three flux ratios are plotted where $ f_i $ refers to the flux of the $ i $th lensed image. The blue bins depict the distributions of subhalos from the empirical model, the orange bins show the subhalo distributions for the emulator model, and the red bins show flux ratios when the lensing galaxy contains no subhalos.}
    \label{flux_ratios}
\end{figure*}

As an example of emulated subhalos being applied in a lensing analysis, we construct $ S_\text{lens} $ distributions from 10,000 realizations of subhalo populations from both empirical and emulated subhalo models. Figure \ref{flux_ratios} shows the distributions of flux ratios between the empirical and emulator models. We see that there is very good agreement between the distribution of flux ratios, indicating that the emulator produces subhalo populations with lensing properties comparable to those in the empirical model. Additionally, the red bins show the case where there is no dark matter substructure. We see that in the $ f_2/f_1 $ and $ f_3/f_1 $ ratios that there are noticeable differences between the red bins and the other two models. 

\begin{figure*}
    \includegraphics[width = \linewidth]{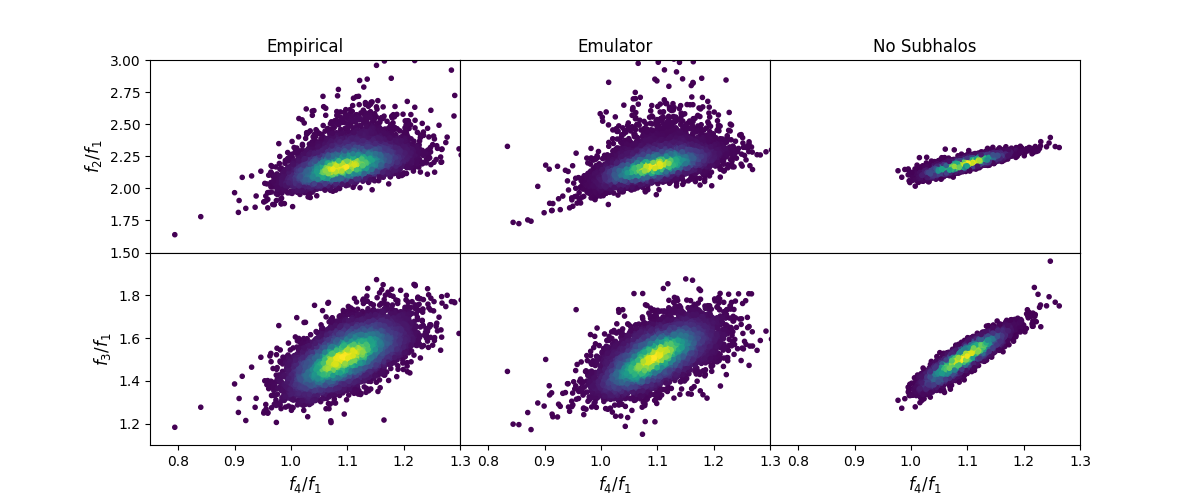}
    \centering
    \caption{Two dimensional density distributions of flux ratios for the empirical model (left column), emulator model (middle column) and model without substructure (right column).}
    \label{flux_densities}
\end{figure*}

Looking at the $ f_4/f_1 $ ratios however, it appears that the distribution of flux ratios is nearly identical, regardless of if the model contains substructure or not. We investigate this further in Figure \ref{flux_densities}, which shows two dimensional distributions of the flux ratios. We see that the density distributions for the no subhalos model appears noticeably different from both the empirical and emulator models. This suggests that the emulator reproduces flux ratios more closely aligned with the empirical model than with a model devoid of dark matter subhalos. Additionally, this suggests that the similar $ f_4/f_1 $ histograms in Figure \ref{flux_ratios} are simply a consequence of marginalization, and do not imply that the underlying flux ratio distributions are nearly identical.

\begin{figure*}
    \includegraphics[width = \linewidth]{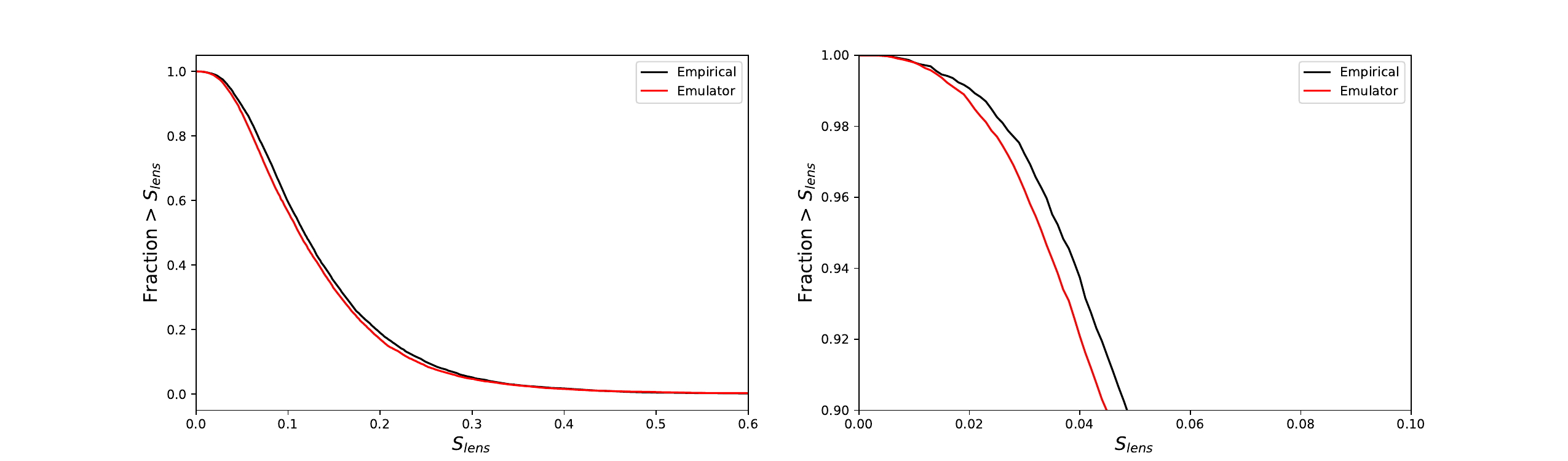}
    \centering
    \caption{\emph{Left panel:} inverted CDFs of subhalo populations from both the empirical model (black) and the emulator model (red). \emph{Right panel:} a zoom-in on the left plot for the lowest $ S_\text{lens} $ values.}
    \label{s_lens}
\end{figure*}

Figure \ref{s_lens} shows the cumulative distribution of $ S_\text{lens} $ values from both models. The plot on the left shows the cumulative distribution function (CDF) of the $ S_\text{lens} $ summary statistic. The plot on the right shows a zoomed-in version of the CDFs for $ S_\text{lens} \leq 0.1 $. Subhalo realizations in the right-hand plot are especially of interest, as these populations produce flux ratios that most closely resemble the target data. We see that emulated subhalos when implemented into the forward modeling analysis, produce comparable results to the empirical model. This is expected (assuming that our emulation procedure accurately captures the distribution of subhalo properties predicted by {\sc Galacticus}) as the empirical model itself was constructed to approximately match the results of {\sc Galacticus} simulations.

One notable difference between the empirical and emulator models is the average number of subhalos per realization. On average, the empirical model produces around 23,000 subhalos per realization, compared to only 1,200 subhalos per realization coming from the emulator. The reason for this difference is the choice of how the subhalo mass resolution is implemented in each model. For {\sc Galacticus}, the mass resolution is applied to the bound mass of subhalos---that is, any subhalo that falls below a bound mass of $10^6\mathrm{M}_\odot$ is removed from the population\footnote{This choice is made for computational efficiency and because, even if such subhalos were retained, the population of subhalos with $M_\mathrm{bound} < 10^6\mathrm{M}_\odot$ would be incomplete due to the missing contribution from subhalos with $M_\mathrm{infall} < 10^6\mathrm{M}_\odot$.}. Conversely, in the empirical model, the mass resolution is applied to infall masses, such that a subhalo may have a bound mass much smaller than $10^6\mathrm{M}_\odot$---such halos are \emph{not} discarded in the empirical model. We find that $\sim$90\% of empirical subhalos have $M_\mathrm{bound} < 10^6\mathrm{M}_\odot$, which is to be expected \citep{2025arXiv250307728D}.

\begin{figure}
    \includegraphics[width = \linewidth]{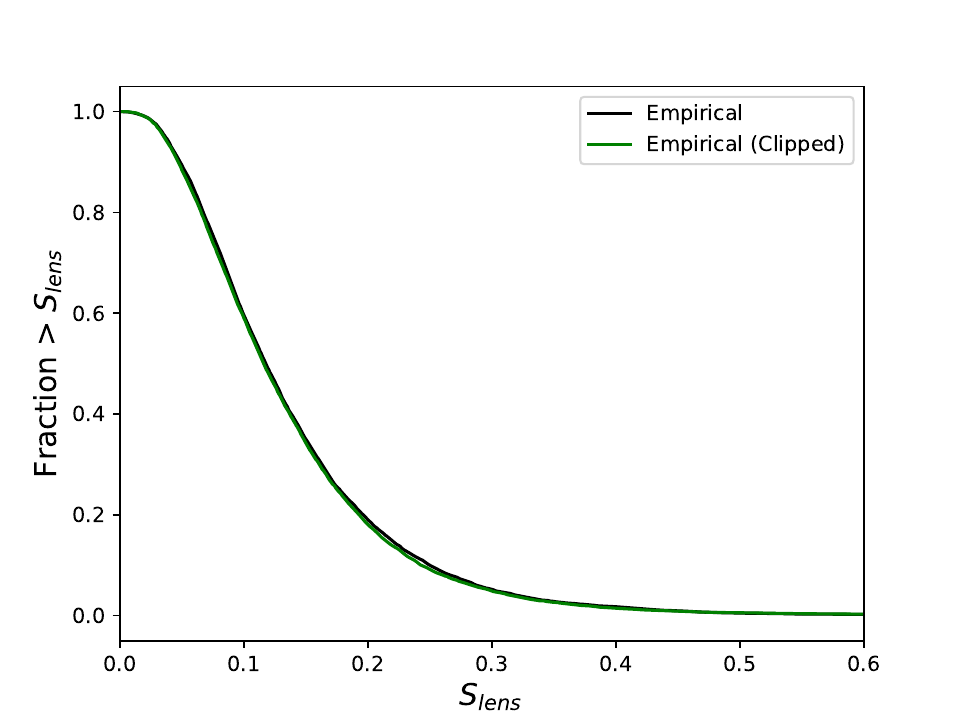}
    \centering
    \caption{Distribution of $ S_\text{lens} $ values for the empirical model. The black curve includes subhalos which have infall masses above $ 10^6 \mathrm{M}_\odot $ and bound masses below $ 10^6 \mathrm{M}_\odot $. The green curve excludes these halos.}
    \label{clipped}
\end{figure}

A natural question to ask is whether or not the additional subhalos from the empirical model that are excluded in the emulator model have a significant impact on the resulting $ S_\text{lens} $ distribution curves. In Figure \ref{clipped}, we plot the $ S_\text{lens} $ distributions for the empirical model in two cases: One instance where we include subhalos with infall masses above $ 10^6 \mathrm{M}_\odot $ but bound masses below $ 10^6 \mathrm{M}_\odot $ (black), and one instance where we exclude these halos (green). By excluding subhalos whose bound mass falls below the resolution, we remove the vast majority of subhalos from a given population. However despite removing such a large portion of the subhalo population, the distribution curves are essentially overlapping. This further supports the work from \cite{gilman2020warm}, which shows that these low-mass subhalos do not significantly affect the observed lensing signal.

\section{Limitations/Future Work} \label{sec:limitations}

Here, we highlight the limitations of our emulator approach as well as possibilities for future work. One limitation of the emulator approach arises from how the normalizing flow is constructed. Requiring that the map between latent and data spaces be differentiable and invertible restricts the possible forms that the map can take, and thus certain distributions become difficult to replicate (i.e. distributions with sharp edges, multimodal distributions, regions with complex topologies, etc.). An example of this limitation is discussed in Figure \ref{con_cdf} in Appendix \ref{sec:appA}. Another limitation of this current approach is that the emulator knows nothing about sub-subhalos, as this information is not provided to it during training, so it is unable to capture correlations between these halos and their host subhalos. While this is expected to have only a small effect on the resulting subhalo populations (sub-subhalos typically make up less than 10\% of the total population in {\sc Galacticus}), future work could avoid this issue by using a second emulator to populate each subhalo with sub-subhalos for example.

There are several other directions that could be explored in future work to improve upon the results of this study. For example, in this work, we used the emulator to generate a six-dimensional joint distribution of subhalo parameters for a specific host halo mass and redshift and for the case of cold dark matter. Previous studies have demonstrated how normalizing flow algorithms can be used to learn conditional distributions \citep{winkler2019learning, friedman2022higlow, abbasi2023conditional, nguyen2024dreams}{---}often referred to as conditional normalizing flow (CNF) algorithms.

By implementing a CNF algorithm we could greatly increase the flexibility of this model. This generalization can, in principle, be applied to any set of interdependent parameters, but two specific areas of interest are as follows. First, the emulator could be conditioned on host halo mass and redshift, allowing a single emulator to be trained and applied to all observed lenses. Another potentially powerful application of a CNF would be to construct an emulator conditioned on the physics of a given dark matter model. For example, performing the same lensing analysis for a warm dark matter (WDM) model would require knowledge of the WDM particle mass. Current studies estimating $ m_\text{WDM} \gtrsim 7 $ keV \citep{nadler2021dark, nierenberg2023jwst, keeley2024jwst}. With our current approach, one would need to train a separate emulator for each WDM particle mass to be considered. If a CNF algorithm was implemented, however, the same 6D parameter space could be learned as a function of WDM particle mass, allowing rapid exploration of WDM models, and for constraints on WDM particle mass to be derived. This would, of course, require the generation of {\sc Galacticus} realization of subhalo populations for WDM models of varying masses in order to train the CNF. Generalization to other, beyond-CDM dark matter models (e.g., conditioning on any self-interaction cross-section of the dark matter particle) may also be possible.

\section{Conclusions} \label{sec:conclusions}

In this work, we examined the strong gravitational lensing perturbations to flux ratios of a quadruply lensed quasar resulting from a population of subhalos produced by the empirical model of \cite{gilman2020warm} as updated by \cite{2024MNRAS.535.1652K}, and compared the results against same analysis conducted using subhalo population generated from {\sc Galacticus} via a normalizing flow emulator. A statistical approach was used in which many subhalo populations were generated, but we only kept the realizations that most closely resembled observational data for both the empirical and emulator models. This emulation approach makes such analyses computationally feasible---generating sufficient realizations directly from {\sc Galacticus} would be computationally impractical.

We found that the normalizing flows algorithm accurately replicates subhalo populations, achieving this in a fraction of the time required for direct generation from {\sc Galacticus}. The distributions of summary statistics describing the perturbations are similar (although not identical) from the empirical model and emulated {\sc Galacticus}---this is as expected as the empirical model was constructed to approximately match {\sc Galacticus} simulations, and serves as a demonstration that results derived using the empirical model are not substantially affected by the simplifications made in that approach. 

In future work, we aim to extend this approach by generalizing the normalizing flows algorithm used to generate subhalo populations. We currently train the algorithm to learn a joint distribution of the subhalo parameter space. In future work, we aim to use the emulator to learn a conditional distribution, allowing subhalo populations to be generated as a function of, for example, the mass of a warm dark matter particle. Doing this will allow us to test the strong lensing signature for many different WDM models, and thereby place accurate constraints on the mass of a WDM particle.

\section*{Acknowledgments}

We thank Charles Gannon for their helpful discussions and their assistance in calibrating the empirical model. We also thank the anonymous referee for comments and suggestions that have improved this paper.

\bibliographystyle{aasjournal}

\bibliography{main}

\begin{thebibliography}{}
\expandafter\ifx\csname natexlab\endcsname\relax\def\natexlab#1{#1}\fi
\providecommand{\url}[1]{\href{#1}{#1}}
\providecommand{\dodoi}[1]{doi:~\href{http://doi.org/#1}{\nolinkurl{#1}}}
\providecommand{\doeprint}[1]{\href{http://ascl.net/#1}{\nolinkurl{http://ascl.net/#1}}}
\providecommand{\doarXiv}[1]{\href{https://arxiv.org/abs/#1}{\nolinkurl{https://arxiv.org/abs/#1}}}

\bibitem[{Abbasi {et~al.}(2023)Abbasi, Ackermann, Adams, Agarwalla, Aguilar, Ahlers, Alameddine, Amin, Andeen, \& Anton}]{abbasi2023conditional}
Abbasi, R., Ackermann, M., Adams, J., {et~al.} 2023, Conditional normalizing flows for IceCube event reconstruction, Tech. rep.

\bibitem[{Ade {et~al.}(2016)Ade, Aghanim, Arnaud, Ashdown, Aumont, Baccigalupi, Banday, Barreiro, Bartlett, Bartolo, {et~al.}}]{ade2016planck}
Ade, P.~A., Aghanim, N., Arnaud, M., {et~al.} 2016, Astronomy \& Astrophysics, 594, A13

\bibitem[{Aghanim {et~al.}(2020)Aghanim, Akrami, Ashdown, Aumont, Baccigalupi, Ballardini, Banday, Barreiro, Bartolo, Basak, {et~al.}}]{aghanim2020planck}
Aghanim, N., Akrami, Y., Ashdown, M., {et~al.} 2020, Astronomy \& Astrophysics, 641, A6

\bibitem[{Bagla(2005)}]{bagla2005cosmological}
Bagla, J.~S. 2005, Current science, 1088

\bibitem[{Benson(2012)}]{benson2012galacticus}
Benson, A.~J. 2012, New Astronomy, 17, 175

\bibitem[{Benson \& Du(2022)}]{benson2022tidal}
Benson, A.~J., \& Du, X. 2022, Monthly Notices of the Royal Astronomical Society, 517, 1398

\bibitem[{{Benson} {et~al.}(2004){Benson}, {Lacey}, {Frenk}, {Baugh}, \& {Cole}}]{2004MNRAS.351.1215B}
{Benson}, A.~J., {Lacey}, C.~G., {Frenk}, C.~S., {Baugh}, C.~M., \& {Cole}, S. 2004, \mnras, 351, 1215, \dodoi{10.1111/j.1365-2966.2004.07870.x}

\bibitem[{Bertschinger(1998)}]{bertschinger1998simulations}
Bertschinger, E. 1998, Annual Review of Astronomy and Astrophysics, 36, 599

\bibitem[{{Bower}(1991)}]{1991MNRAS.248..332B}
{Bower}, R.~G. 1991, \mnras, 248, 332, \dodoi{10.1093/mnras/248.2.332}

\bibitem[{Bryan \& Norman(1998)}]{bryan1998statistical}
Bryan, G.~L., \& Norman, M.~L. 1998, The Astrophysical Journal, 495, 80

\bibitem[{Bullock \& Boylan-Kolchin(2017)}]{bullock2017small}
Bullock, J.~S., \& Boylan-Kolchin, M. 2017, Annual Review of Astronomy and Astrophysics, 55, 343

\bibitem[{Cole \& Lacey(1996)}]{cole1996structure}
Cole, S., \& Lacey, C. 1996, Monthly Notices of the Royal Astronomical Society, 281, 716

\bibitem[{Cole {et~al.}(2000)Cole, Lacey, Baugh, \& Frenk}]{cole2000hierarchical}
Cole, S., Lacey, C.~G., Baugh, C.~M., \& Frenk, C.~S. 2000, Monthly Notices of the Royal Astronomical Society, 319, 168

\bibitem[{Dalal \& Kochanek(2002)}]{dalal2002direct}
Dalal, N., \& Kochanek, C. 2002, The Astrophysical Journal, 572, 25

\bibitem[{{Davis} {et~al.}(1985){Davis}, {Efstathiou}, {Frenk}, \& {White}}]{1985ApJ...292..371D}
{Davis}, M., {Efstathiou}, G., {Frenk}, C.~S., \& {White}, S.~D.~M. 1985, \apj, 292, 371, \dodoi{10.1086/163168}

\bibitem[{Dehnen \& Read(2011)}]{dehnen2011n}
Dehnen, W., \& Read, J.~I. 2011, The European Physical Journal Plus, 126, 55

\bibitem[{{Dekker} {et~al.}(2022){Dekker}, {Ando}, {Correa}, \& {Ng}}]{2022PhRvD.106l3026D}
{Dekker}, A., {Ando}, S., {Correa}, C.~A., \& {Ng}, K. C.~Y. 2022, \prd, 106, 123026, \dodoi{10.1103/PhysRevD.106.123026}

\bibitem[{Diemer \& Joyce(2019)}]{diemer2019accurate}
Diemer, B., \& Joyce, M. 2019, The Astrophysical Journal, 871, 168

\bibitem[{{Du} {et~al.}(2025){Du}, {Gilman}, {Treu}, {Benson}, \& {Gannon}}]{2025arXiv250307728D}
{Du}, X., {Gilman}, D., {Treu}, T., {Benson}, A., \& {Gannon}, C. 2025, arXiv e-prints, arXiv:2503.07728, \dodoi{10.48550/arXiv.2503.07728}

\bibitem[{{Dutton} {et~al.}(2016){Dutton}, {Macci{\`o}}, {Frings}, {Wang}, {Stinson}, {Penzo}, \& {Kang}}]{2016MNRAS.457L..74D}
{Dutton}, A.~A., {Macci{\`o}}, A.~V., {Frings}, J., {et~al.} 2016, \mnras, 457, L74, \dodoi{10.1093/mnrasl/slv193}

\bibitem[{{Fitts} {et~al.}(2019){Fitts}, {Boylan-Kolchin}, {Bozek}, {Bullock}, {Graus}, {Robles}, {Hopkins}, {El-Badry}, {Garrison-Kimmel}, {Faucher-Gigu{\`e}re}, {Wetzel}, \& {Kere{\v{s}}}}]{2019MNRAS.490..962F}
{Fitts}, A., {Boylan-Kolchin}, M., {Bozek}, B., {et~al.} 2019, \mnras, 490, 962, \dodoi{10.1093/mnras/stz2613}

\bibitem[{Frenk {et~al.}(1985)Frenk, White, Efstathiou, \& Davis}]{frenk1985cold}
Frenk, C.~S., White, S.~D., Efstathiou, G., \& Davis, M. 1985, Nature, 317, 595

\bibitem[{Friedman \& Hassan(2022)}]{friedman2022higlow}
Friedman, R., \& Hassan, S. 2022, arXiv preprint arXiv:2211.12724

\bibitem[{Gannon {et~al.}(2025)Gannon, Nierenberg, Benson, Keeley, Du, \& Gilman}]{gannon2025dark}
Gannon, C., Nierenberg, A., Benson, A., {et~al.} 2025, arXiv preprint arXiv:2501.17362

\bibitem[{Gilman {et~al.}(2024)Gilman, Birrer, Nierenberg, \& Oh}]{gilman2024turbocharging}
Gilman, D., Birrer, S., Nierenberg, A., \& Oh, M.~S. 2024, Monthly Notices of the Royal Astronomical Society, 533, 1687

\bibitem[{Gilman {et~al.}(2020)Gilman, Birrer, Nierenberg, Treu, Du, \& Benson}]{gilman2020warm}
Gilman, D., Birrer, S., Nierenberg, A., {et~al.} 2020, Monthly Notices of the Royal Astronomical Society, 491, 6077

\bibitem[{Gilman {et~al.}(2021)Gilman, Bovy, Treu, Nierenberg, Birrer, Benson, \& Sameie}]{gilman2021strong}
Gilman, D., Bovy, J., Treu, T., {et~al.} 2021, Monthly Notices of the Royal Astronomical Society, 507, 2432

\bibitem[{{Giocoli} {et~al.}(2008){Giocoli}, {Pieri}, \& {Tormen}}]{2008MNRAS.387..689G}
{Giocoli}, C., {Pieri}, L., \& {Tormen}, G. 2008, \mnras, 387, 689, \dodoi{10.1111/j.1365-2966.2008.13283.x}

\bibitem[{Han {et~al.}(2016)Han, Cole, Frenk, \& Jing}]{han2016unified}
Han, J., Cole, S., Frenk, C.~S., \& Jing, Y. 2016, Monthly Notices of the Royal Astronomical Society, 457, 1208

\bibitem[{Hezaveh {et~al.}(2016)Hezaveh, Dalal, Marrone, Mao, Morningstar, Wen, Blandford, Carlstrom, Fassnacht, Holder, {et~al.}}]{hezaveh2016detection}
Hezaveh, Y.~D., Dalal, N., Marrone, D.~P., {et~al.} 2016, The Astrophysical Journal, 823, 37

\bibitem[{Hinshaw {et~al.}(2013)Hinshaw, Larson, Komatsu, Spergel, Bennett, Dunkley, Nolta, Halpern, Hill, Odegard, {et~al.}}]{hinshaw2013nine}
Hinshaw, G., Larson, D., Komatsu, E., {et~al.} 2013, The Astrophysical Journal Supplement Series, 208, 19

\bibitem[{Hsueh {et~al.}(2020)Hsueh, Enzi, Vegetti, Auger, Fassnacht, Despali, Koopmans, \& McKean}]{hsueh2020sharp}
Hsueh, J.-W., Enzi, W., Vegetti, S., {et~al.} 2020, Monthly Notices of the Royal Astronomical Society, 492, 3047

\bibitem[{Jiang \& van~den Bosch(2017)}]{jiang2017statistics}
Jiang, F., \& van~den Bosch, F.~C. 2017, Monthly Notices of the Royal Astronomical Society, 472, 657

\bibitem[{{Keeley} {et~al.}(2024){Keeley}, {Nierenberg}, {Gilman}, {Gannon}, {Birrer}, {Treu}, {Benson}, {Du}, {Abazajian}, {Anguita}, {Bennert}, {Djorgovski}, {Gupta}, {Hoenig}, {Kusenko}, {Lemon}, {Malkan}, {Motta}, {Moustakas}, {Oh}, {Sluse}, {Stern}, \& {Wechsler}}]{2024MNRAS.535.1652K}
{Keeley}, R.~E., {Nierenberg}, A.~M., {Gilman}, D., {et~al.} 2024, \mnras, 535, 1652, \dodoi{10.1093/mnras/stae2458}

\bibitem[{Keeley {et~al.}(2024)Keeley, Nierenberg, Gilman, Gannon, Birrer, Treu, Benson, Du, Abazajian, Anguita, {et~al.}}]{keeley2024jwst}
Keeley, R.~E., Nierenberg, A.~M., Gilman, D., {et~al.} 2024, arXiv preprint arXiv:2405.01620

\bibitem[{{Lacey} \& {Cole}(1994)}]{1994MNRAS.271..676L}
{Lacey}, C., \& {Cole}, S. 1994, \mnras, 271, 676, \dodoi{10.1093/mnras/271.3.676}

\bibitem[{Loshchilov(2017)}]{loshchilov2017decoupled}
Loshchilov, I. 2017, arXiv preprint arXiv:1711.05101

\bibitem[{Makishima(1998)}]{makishima1998hierarchical}
Makishima, K. 1998, in Symposium-International Astronomical Union, Vol. 188, Cambridge University Press, 181--184

\bibitem[{Mikkola \& Aarseth(1993)}]{mikkola1993implementation}
Mikkola, S., \& Aarseth, S.~J. 1993, Celestial Mechanics and Dynamical Astronomy, 57, 439

\bibitem[{{Moore} {et~al.}(1999){Moore}, {Ghigna}, {Governato}, {Lake}, {Quinn}, {Stadel}, \& {Tozzi}}]{1999ApJ...524L..19M}
{Moore}, B., {Ghigna}, S., {Governato}, F., {et~al.} 1999, \apjl, 524, L19, \dodoi{10.1086/312287}

\bibitem[{Nadler {et~al.}(2021)Nadler, Birrer, Gilman, Wechsler, Du, Benson, Nierenberg, \& Treu}]{nadler2021dark}
Nadler, E.~O., Birrer, S., Gilman, D., {et~al.} 2021, The Astrophysical Journal, 917, 7

\bibitem[{Nadler {et~al.}(2023)Nadler, Mansfield, Wang, Du, Adhikari, Banerjee, Benson, Darragh-Ford, Mao, Wagner-Carena, {et~al.}}]{nadler2023symphony}
Nadler, E.~O., Mansfield, P., Wang, Y., {et~al.} 2023, The Astrophysical Journal, 945, 159

\bibitem[{Navarro(1996)}]{navarro1996structure}
Navarro, J.~F. 1996, in Symposium-international astronomical union, Vol. 171, Cambridge University Press, 255--258

\bibitem[{Nguyen {et~al.}(2024)Nguyen, Villaescusa-Navarro, Mishra-Sharma, Cuesta-Lazaro, Torrey, Farahi, Garcia, Rose, O'Neil, Vogelsberger, {et~al.}}]{nguyen2024dreams}
Nguyen, T., Villaescusa-Navarro, F., Mishra-Sharma, S., {et~al.} 2024, arXiv preprint arXiv:2409.02980

\bibitem[{Nierenberg {et~al.}(2014)Nierenberg, Treu, Wright, Fassnacht, \& Auger}]{nierenberg2014detection}
Nierenberg, A., Treu, T., Wright, S., Fassnacht, C., \& Auger, M. 2014, Monthly Notices of the Royal Astronomical Society, 442, 2434

\bibitem[{Nierenberg {et~al.}(2023)Nierenberg, Keeley, Sluse, Gilman, Birrer, Treu, Abazajian, Anguita, Benson, Bennert, {et~al.}}]{nierenberg2023jwst}
Nierenberg, A., Keeley, R., Sluse, D., {et~al.} 2023, arXiv preprint arXiv:2309.10101

\bibitem[{Papamakarios {et~al.}(2021)Papamakarios, Nalisnick, Rezende, Mohamed, \& Lakshminarayanan}]{papamakarios2021normalizing}
Papamakarios, G., Nalisnick, E., Rezende, D.~J., Mohamed, S., \& Lakshminarayanan, B. 2021, Journal of Machine Learning Research, 22, 1

\bibitem[{Parkinson {et~al.}(2008)Parkinson, Cole, \& Helly}]{parkinson2008generating}
Parkinson, H., Cole, S., \& Helly, J. 2008, Monthly Notices of the Royal Astronomical Society, 383, 557

\bibitem[{Peebles(2020)}]{peebles2020principles}
Peebles, P. J.~E. 2020, Principles of physical cosmology (Princeton university press)

\bibitem[{P{\'e}rez-Cruz(2008)}]{perez2008kullback}
P{\'e}rez-Cruz, F. 2008, in 2008 IEEE international symposium on information theory, IEEE, 1666--1670

\bibitem[{Pullen {et~al.}(2014)Pullen, Benson, \& Moustakas}]{pullen2014nonlinear}
Pullen, A.~R., Benson, A.~J., \& Moustakas, L.~A. 2014, The Astrophysical Journal, 792, 24

\bibitem[{{Sheth} \& {Tormen}(2002)}]{2002MNRAS.329...61S}
{Sheth}, R.~K., \& {Tormen}, G. 2002, \mnras, 329, 61, \dodoi{10.1046/j.1365-8711.2002.04950.x}

\bibitem[{Spergel {et~al.}(2003)Spergel, Verde, Peiris, Komatsu, Nolta, Bennett, Halpern, Hinshaw, Jarosik, Kogut, {et~al.}}]{spergel2003first}
Spergel, D.~N., Verde, L., Peiris, H.~V., {et~al.} 2003, The Astrophysical Journal Supplement Series, 148, 175

\bibitem[{{Springel} {et~al.}(2008){Springel}, {Wang}, {Vogelsberger}, {Ludlow}, {Jenkins}, {Helmi}, {Navarro}, {Frenk}, \& {White}}]{2008MNRAS.391.1685S}
{Springel}, V., {Wang}, J., {Vogelsberger}, M., {et~al.} 2008, \mnras, 391, 1685, \dodoi{10.1111/j.1365-2966.2008.14066.x}

\bibitem[{{Taylor} \& {Babul}(2001)}]{2001ApJ...559..716T}
{Taylor}, J.~E., \& {Babul}, A. 2001, \apj, 559, 716, \dodoi{10.1086/322276}

\bibitem[{Treu(2010)}]{treu2010strong}
Treu, T. 2010, Annual Review of Astronomy and Astrophysics, 48, 87

\bibitem[{Vegetti \& Vogelsberger(2014)}]{vegetti2014density}
Vegetti, S., \& Vogelsberger, M. 2014, Monthly Notices of the Royal Astronomical Society, 442, 3598

\bibitem[{Winkler {et~al.}(2019)Winkler, Worrall, Hoogeboom, \& Welling}]{winkler2019learning}
Winkler, C., Worrall, D., Hoogeboom, E., \& Welling, M. 2019, arXiv preprint arXiv:1912.00042

\bibitem[{Xu {et~al.}(2015)Xu, Sluse, Gao, Wang, Frenk, Mao, Schneider, \& Springel}]{xu2015well}
Xu, D., Sluse, D., Gao, L., {et~al.} 2015, Monthly Notices of the Royal Astronomical Society, 447, 3189

\bibitem[{Yang {et~al.}(2020)Yang, Du, Benson, Pullen, \& Peter}]{yang2020new}
Yang, S., Du, X., Benson, A.~J., Pullen, A.~R., \& Peter, A.~H. 2020, Monthly Notices of the Royal Astronomical Society, 498, 3902

\bibitem[{{Yang} {et~al.}(2011){Yang}, {Mo}, {Zhang}, \& {van den Bosch}}]{2011ApJ...741...13Y}
{Yang}, X., {Mo}, H.~J., {Zhang}, Y., \& {van den Bosch}, F.~C. 2011, \apj, 741, 13, \dodoi{10.1088/0004-637X/741/1/13}

\bibitem[{{Zentner} {et~al.}(2005){Zentner}, {Berlind}, {Bullock}, {Kravtsov}, \& {Wechsler}}]{2005ApJ...624..505Z}
{Zentner}, A.~R., {Berlind}, A.~A., {Bullock}, J.~S., {Kravtsov}, A.~V., \& {Wechsler}, R.~H. 2005, \apj, 624, 505, \dodoi{10.1086/428898}

\end{thebibliography}

\begin{appendix}

\section{Accuracy of the Emulator} \label{sec:appA}

Here, we describe tests that quantify the accuracy of the emulator in its ability to replicate {\sc Galacticus} subhalo populations. We first conducted a two-sample Kolmogorov-Smirnov (KS) test, where a test statistic $ D_{m,n} $ is defined by:
\begin{equation}
    D_{m,n} = \sup_x \left| F_m(x) - G_n(x) \right|,
\end{equation}
where $ m, n $ are the sizes of the two data sets, and $ D_{m,n} $ represents the maximum absolute difference between two distributions. Table \ref{2SKS} shows a summary of the $ D_{m,n} $ statistics comparing the {\sc Galacticus} and emulator (trained on {\sc Galacticus}) distributions:

\begin{table}[hbt!]
    \centering
    \caption{Results from the two-sample KS test.}
        \begin{tabular}{ll}
            \hline
            \textbf{Parameter} & \boldmath{$ D_{m,n} $} \\
            \hline
            Infall mass & 0.008 \\
            Concentration & 0.023 \\
            Bound mass & 0.016 \\
            Infall redshift & 0.008 \\
            Truncation radius & 0.013 \\
            2D projected radius & 0.009 \\
            \hline
        \end{tabular}
    \label{2SKS}
\end{table}

The $ D_{m,n} $ values from Table \ref{2SKS} imply that the {\sc Galacticus} and emulator distributions are not identical. While this may seem problematic, the goal is not necessarily for the emulator to exactly replicate the distribution of {\sc Galacticus} subhalos. Rather, the goal is for the emulator to produce a strong lensing signal that is sufficiently close to the strong lensing signature that would have been produced by {\sc Galacticus} subhalos. Here, ``sufficiently close'' means that remaining differences should be small compared to other approximations made in the {\sc Galacticus} modeling, which are at least of order 10\% \citep{nadler2023symphony}.

\begin{figure}
    \includegraphics[width = \linewidth]{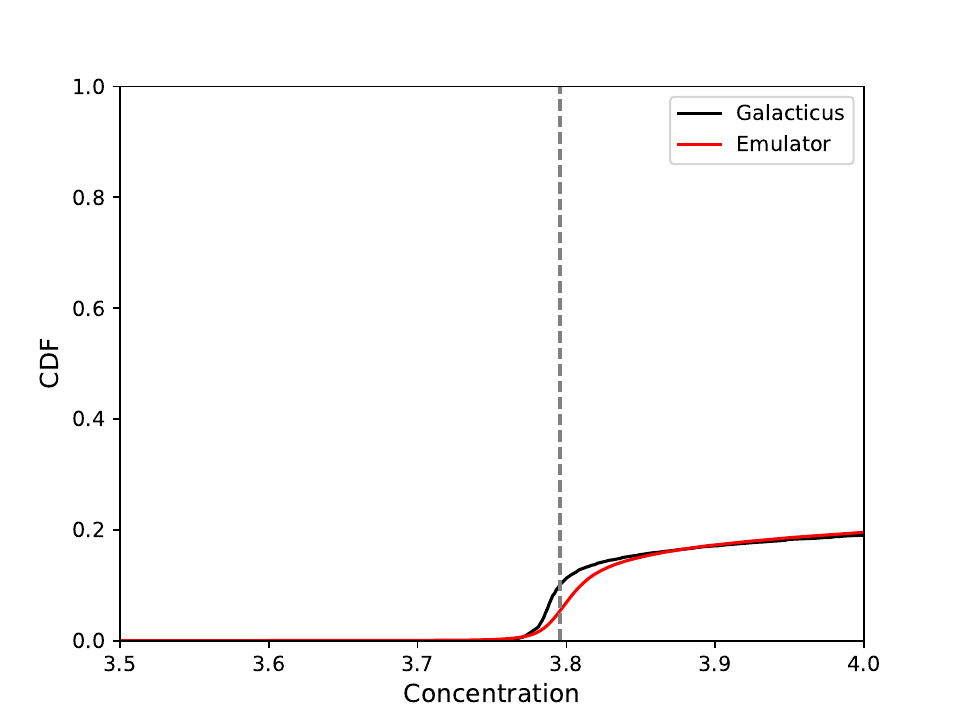}
    \centering
    \caption{CDFs of the concentration parameter for both the {\sc Galacticus} and emulator distributions, zoomed in to the region around the minimum allowed concentration in {\sc Galacticus}, which is where the maximum discrepancy occurs between distributions. The vertical dashed line represents the value of $ c $ at which the maximum discrepancy between distributions occurs.}
    \label{con_cdf}
\end{figure}

We investigate the source of the discrepancy between distributions by comparing their CDFs. For example, Figure \ref{con_cdf} shows the CDF of the concentration parameter, the parameter with the largest $ D_{m,n} $ value. Figure~\ref{con_cdf} compares CDFs of the concentration parameter $ c $ between the {\sc Galacticus} (black) and emulator (red) subhalo distributions, and is zoomed in on a region around $ c = 3.795 $  (grey dashed line) where $ D_{m,n} $ is maximized. There is a relatively steep increase in the number of {\sc Galacticus} subhalos around $ c = 3.795 $, which is the minimum concentration of subhalos in the simulation\footnote{{\sc Galacticus} imposes a minimum concentration of $c_\mathrm{vir}=4$, defined relative to the spherical collapse virial radius. Here we show concentration, $c$, defined relative to the $R_{200}$ definition of virial radius, which results in a small, redshift-dependent shift in the cut-off, causing the cut-off to be somewhat smoothed out.}. The emulator CDF does not replicate this sharp increase perfectly, which can be attributed to the inherent limitations of a normalizing flows algorithm. Such models tend to have difficulties accurately replicating sharp changes in distribution due to the requirements of invertibility and differentiability of the mapping between latent and data spaces.

The two-sample KS test is intended to test the similarity between two univariate distributions. To examine higher dimensional correlations between the {\sc Galacticus} and emulator distributions, we first generate correlation matrices for normalized {\sc Galacticus} and emulator subhalo parameters, which we denote as $ R_G $ and $ R_E $ respectively.

\begin{table}[hbt!]
    \centering
    \caption{Correlation matrix of normalized {\sc Galacticus} and emulator subhalo parameters}
        \begin{tabular}{l|llllll}
            & $ y_1 $ & $ y_2 $ & $ y_3 $ & $ y_4 $ & $ y_5 $ & $ y_6 $ \\
            \hline
            $ y_1 $ & 1.000 & -0.501 & -0.796 & 0.381 & -0.396 & 0.003 \\
            $ y_2 $ & -0.494 & 1.000 & 0.616 & -0.704 & 0.699 & -0.006 \\
            $ y_3 $ & -0.795 & 0.613 & 1.000 & -0.562 & 0.802 & 0.003 \\
            $ y_4 $ & 0.370 & -0.695 & -0.551 & 1.000 & -0.760 & 0.001 \\
            $ y_5 $ & -0.392 & 0.698 & 0.804 & -0.760 & 1.000 & 0.001 \\
            $ y_6 $ & 0.002 & -0.030 & -0.026 & 0.029 & -0.040 & 1.000 \\
        \end{tabular}
    \label{RG}
\end{table}

Table \ref{RG} shows the elements of the $ R_G $ and $ R_E $ matrices, where the upper triangular entries represent the correlation coefficients of {\sc Galacticus} data, and the lower triangular entries correspond to emulator data. By comparing off diagonal elements, we see that there are similar linear correlations between the two distributions. For example, the $ y_1 $ versus $ y_3 $ linear correlation for the {\sc Galacticus} distribution is -0.796, whereas the same linear correlation coefficient for the emulator distribution is -0.795. Certain correlation coefficients have larger discrepancies between the two distributions, such as $ y_5 $ versus $ y_6 $. However this can naturally be explained due to noise from sampled data between two uncorrelated variables. To gain an understanding for how similar these linear correlations are collectively, we compute the Frobenius norm for the difference of these matrices $ R \equiv R_G - R_E $.

\begin{equation} \label{norm}
    ||R||_F = \sqrt{\sum_{i,j} |R_{ij}|^2} = 0.103
\end{equation}

To interpret the distance in equation \ref{norm}, we note that the Frobenius norm for the difference in correlation matrices has an upper bound, which would occur if one correlation matrix had $ +1 $ for its off diagonal elements while the other matrix had $ -1 $. In the upper bound case, the Frobenius norm would be $ \sqrt{(6 \cdot 0) + (30 \cdot 4)} = \sqrt{120} \approx 10.95 $. Given that $ 0.103 \ll 10.95 $, this indicates that the linear correlations between {\sc Galacticus} and emulator subhalo populations are similar.

To relate {\sc Galacticus} and emulator distributions beyond their linear dependencies, we also compute the Kullback-Leibler (KL) divergence, $ D_\text{KL} $, between the two distributions. Unlike the Frobenius norm of the difference in correlation matrices, the KL divergence does not have a fixed upper bound. Therefore, to interpret the found  $ D_\text{KL} $ values, we also calculate the KL divergence between the distribution of emulated {\sc Galacticus} subhalos and the distribution of empirical subhalos.

\begin{table}[hbt!]
    \centering
    \caption{KL Divergence estimates comparing emulator distribution to Galacticus and empirical distributions}
        \begin{tabular}{|l|l|}
            \hline
            \textbf{Galacticus} & \textbf{Empirical} \\
            \hline
            $0.213 \pm 0.005$ & $7.511 \pm 0.096$ \\
            \hline
        \end{tabular}
    \label{KL}
\end{table}

Table \ref{KL} summarizes the results for computing the KL divergence between the emulator and {\sc Galacticus} distributions (left column) in addition to the KL divergence between the emulator and empirical distributions (right column). The results from table \ref{KL} used sampled data points to estimate the KL divergence between pairs of continuous distributions. The details for how this estimation is performed can be found in \cite{perez2008kullback}. The KL divergence was computed 5 times with different random samples of points for each pair of distributions to quantify the statistical uncertainty in the obtained $ D_\text{KL} $ values. The KL divergence values imply not only that there are systematic differences between the {\sc Galacticus} emulated and empirical distributions, but also that the emulator is able to capture most of the difference between the two distributions.

\begin{figure}
    \includegraphics[width = \linewidth]{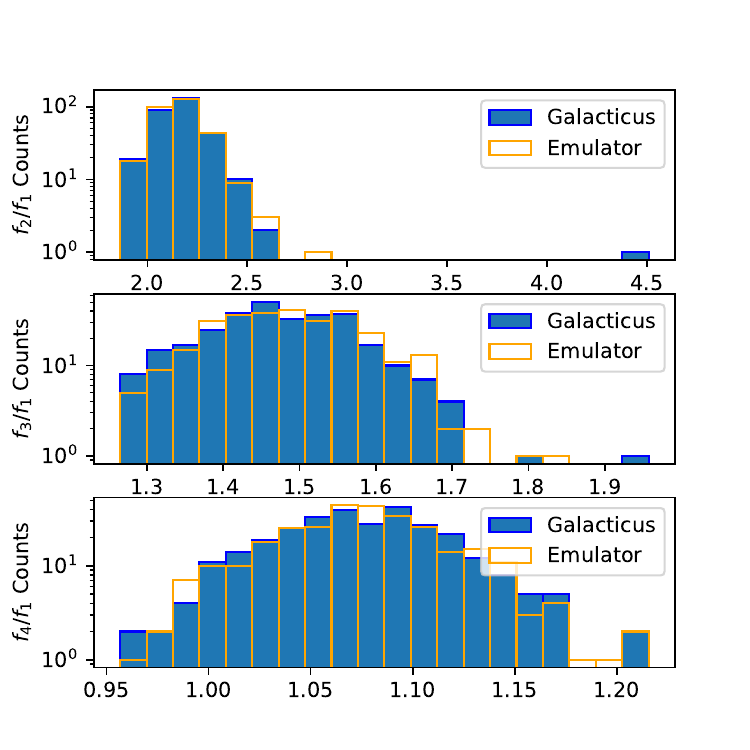}
    \centering
    \caption{Histograms of flux ratios produced from {\sc Galacticus} realizations (blue bins) versus {\sc Galacticus} trained emulator realizations (orange bins). Each set of histograms was composed of 300 subhalo populations.}
    \label{gal_ratios}
\end{figure}

To test if the emulator is able to produce a lensing signal statistically equivalent to the subhalo population it replicates, we run the {\sc Galacticus} realizations used to train the emulator through the forward modeling analysis, and compare the resulting flux ratios. Figure \ref{gal_ratios} shows the distributions for the 300 {\sc Galacticus} realizations (blue bins) against 300 generated emulator subhalo populations (orange bins). We chose to match the number of emulated realizations to the number of {\sc Galacticus} realizations (as opposed to comparing {\sc Galacticus} data against the 10,000 generated realizations from section \ref{sec:r&a}) to obtain a fair comparison of the distributions of flux ratios. A significantly larger number of emulated subhalo realizations increases the number of outlier flux ratio values, which results in a broader spread in the flux ratio distributions, even when normalized. We see that when comparing models with similar amounts of data, we find very similar distributions. This demonstrates that the emulator can accurately reproduce {\sc Galacticus} lensing signatures.

\begin{figure}
    \includegraphics[width = \linewidth]{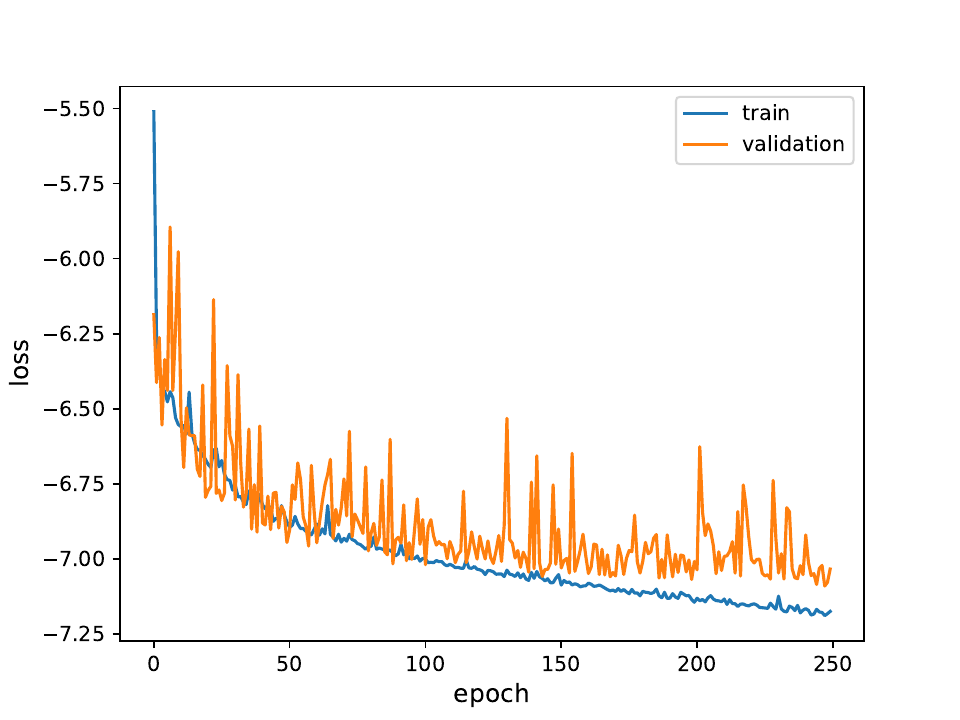}
    \centering
    \caption{The emulator model loss as a function of training epoch. The blue curve shows the model loss for the training dataset, and the orange curve shows the model loss on the validation dataset. }
    \label{loss}
\end{figure}

Lastly, we examine the model loss over the course of the emulator training in Figure~\ref{loss}. The model loss here is the quantity that the algorithm seeks to minimize over the course of the training period to produce an emulator that matches the input data.

In Figure \ref{loss}, the blue and orange curves represent the model loss as a function of the total number of epochs for the training and validation data sets, respectively. There is an 80/20 split across the 300 input {\sc Galacticus} trees between the training and validation sets, respectively. The validation loss starts to flatten out around epochs 50--75, with only a slow decline at further epochs. Nevertheless, the validation model loss \emph{does} still slowly decrease, indicating that our emulator is not strongly affected by overfitting.

\end{appendix}
\end{document}